\documentclass{article}
\usepackage{spconf,amsmath,graphicx,hyperref}
\usepackage{booktabs}
\usepackage[table]{xcolor}
\usepackage{multirow}
\usepackage{array}
\usepackage{pifont} 
\usepackage{fontawesome5}   
\usepackage{xcolor}         
\usepackage{hyperref}       
\usepackage{amssymb,multirow,booktabs}
\usepackage{algorithm}
\usepackage{algpseudocode}
\usepackage{flafter}
\usepackage{threeparttable}
\title{SepRQ : Self-Supervised Speech Mixture Representation Learning \\ via  Mask-Free, Multi-Scale Source Separation}

\name{Severin Baroudi$^{1,3}$,
  Herve Bredin$^{2}$,
  Ricard Marxer$^{1,3}$}
\address{$^{1}$Univ Toulon, Aix Marseille Univ, CNRS, LIS, Toulon, France\\
  $^{2}$pyannoteAI, Toulouse, France \;
  $^{3}$CNRS, ILLS, Montréal, Canada\\
  }
\begin{document}
\ninept
\maketitle
\begin{abstract}

Self-supervised learning (SSL) is standard for speech representation learning, but mainstream models are designed around single-speaker audio, limiting their usefulness in multi-speakers scenarios. We present SepRQ, an open-source SSL framework that replaces masked prediction with a pseudo-source-separation objective over frozen random-projection codebooks. By adopting a novel mask-free, multiresolution approach, SepRQ achieves state-of-the-art performance in Speaker Diarization and Speech Separation on the SUPERB benchmark, surpassing WavLM and other cocktail-party derived SSLs at both Base and Large scales, while requiring only 85.68M inference parameters.  SepRQ also demonstrates strong performance across target-speaker tasks requiring enrollment (such as Target-Speaker Automatic Speech Recognition), and on the challenging multi-domain DIHARD 3 diarization dataset. Notably, we report strong separation capabilities on three-speaker mixtures (WSJ0-3Mix), where current SSL literature struggles. While cocktail-party SSLs remain scarce and closed-source, limited to C-HuBERT and the enrollment-based SA-WavLM, we open-source SepRQ to the community. 
\end{abstract}
\begin{figure*}[t]
  \centering
  \includegraphics[width=\textwidth]{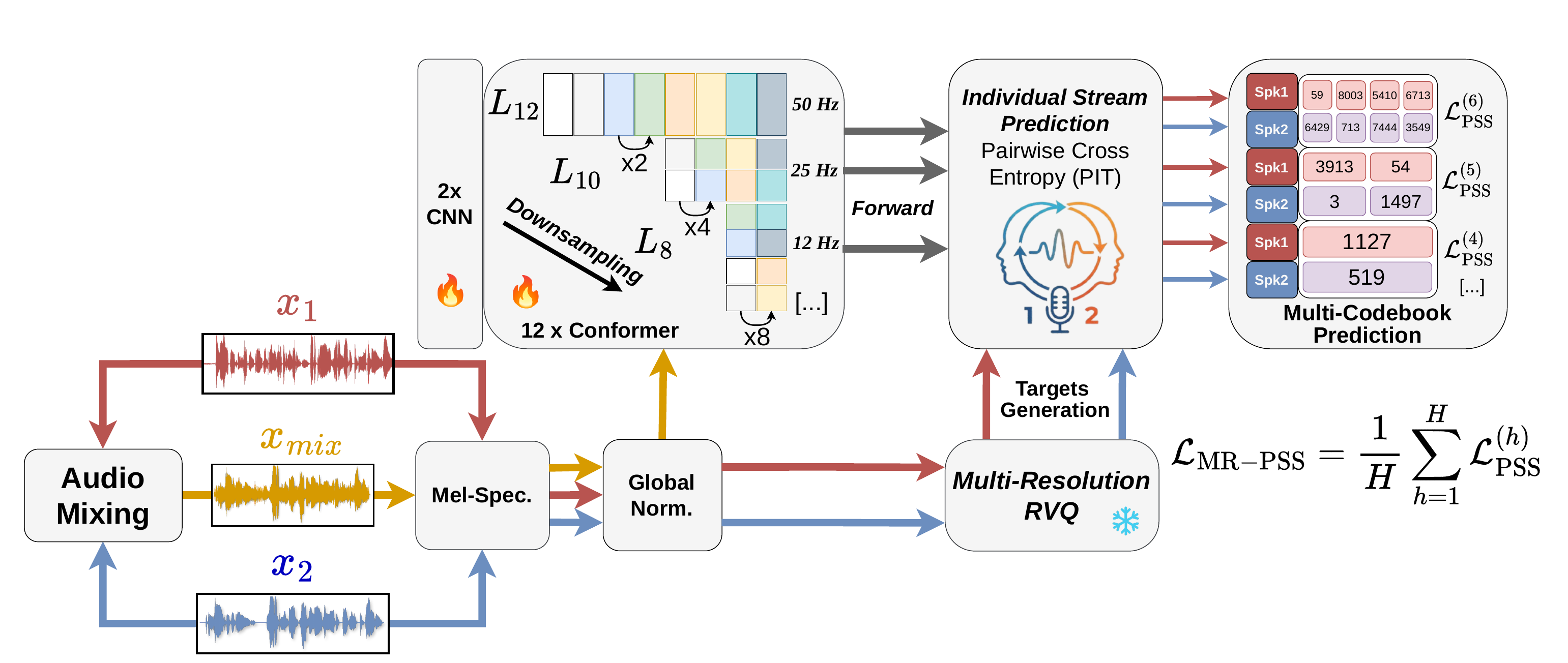}
  \vspace{-2em}
  \caption{Architecture of the proposed SepRQ. The model performs pseudo source separation in the discrete space, using Random Vector Quantizers (RVQs), as speaker-specific discrete labels, in a multiresolution fashion.}
  \label{fig:seprq-multires}
  \vspace{-1em}
\end{figure*}

\section{Introduction}
\label{sec:intro}

Real-world conversational speech often contains overlapping speakers, room reverberation, and background noise, giving rise to the ``cocktail party'' problem of separating and disentangling concurrent speech sources. 
While self-supervised learning (SSL) has been widely used in speech processing, mainstream models such as HuBERT and BestRQ~\cite{wav2vec2,HuBERT,bestrq} have primarily focused on single-speaker tasks, typically using masked prediction on read-speech corpora such as LibriSpeech~\cite{librispeech}. Unlike HuBERT, which relies on an impractical offline $k$-means clustering stage to generate discrete targets before pre-training, BestRQ derives its targets from a single frozen random vector quantizer (RVQ).

Although adapting pretraining data to the target domain strongly influences downstream performance on multi-speaker tasks~\cite{specializing,asru25}, a gap remains in designing SSL architectures explicitly designed for cocktail-party scenarios. The emergence of multi-speaker benchmarks such as TS-SUPERB~\cite{ts_superb} and real-world conversational challenges such as CHiME~\cite{chime6} further highlights this need. Overlapping speech requires representations capable of more than isolated phoneme recognition, since downstream tasks such as Speech Separation (SS), Speaker Diarization (SD), and Target-Speaker Automatic Speech Recognition (TS-ASR) demand disentangling phonetic content, tracking speaker identities, and identifying precise turn boundaries. Although models such as UniSpeech-SAT~\cite{unispeech} or WavLM~\cite{WavLM} improve acoustic robustness through a pretrained denoising task, their core objectives remain inherently single-source. To the best of our knowledge, only few approaches introduce dedicated pretext tasks associated with separation objectives. Cocktail-HuBERT (C-HuBERT)~\cite{cocktail_hubert} enforces both stream separation and discrete masked unit prediction on each target speakers simultaneously, through the use of a Pseudo Source Separation (PSS) objective. On the other hand, Speaker-Aware WavLM (SA-WavLM)~\cite{SA-WAVLM} uses embeddings from enrolled voices to sequentially extract target speakers, relying on necessary single-speaker sources at inference.


Speech tasks differ substantially in the temporal granularity of the information they exploit. ASR performance is sensitive to the temporal resolution of SSL representations, while allowing substantial compression when representations capture appropriate linguistic units~\cite{meng2022compressing}. Speaker diarization presents an explicit multi-timescale trade-off: short segments provide accurate temporal localization, whereas longer segments yield more reliable speaker representations~\cite{park2021multiscale}. Meanwhile, speech separation typically operates on fine-grained acoustic representations to disentangle overlapping sources. These differing temporal requirements motivate our hypothesis that cocktail-party SSL could benefit from representations learned at multiple temporal resolutions. This direction is also consistent with recent multiresolution approaches in single-speaker SSL~\cite{mr_hubert} and supervised speech separation~\cite{multires1,multires2}.
Additionally, while masking can encourage robust representation learning for single-speaker signals, concealing portions of a mixture, as employed by C-HuBERT, may suppress necessary content for speech disentanglement. From an experimental perspective, existing approaches are only evaluated on synthetic mixtures derived only from Librispeech, which differ substantially from real-world speech recordings. Real corpora typically exhibit partial speaker overlap, varying numbers of active speakers, and dynamic turn-taking, making it unclear how well cocktail-party SSL models generalize beyond controlled synthetic conditions. Moreover, C-HuBERT and SA-WavLM remain the only SSL architectures specifically designed for the cocktail-party setting, and the absence of open-source implementations further limits their accessibility and extensibility.

The contributions of this paper are as follows: \textbf{(i)} We introduce SepRQ, an open-source  self-supervised model derived from BestRQ, that features a pseudo source-separation objective targeted on discrete speaker-attributed codebook targets, essentially removing the need for offline clustering employed by WavLM or HuBERT. \textbf{(ii)} We introduce a multiresolution separation objective with progressive downsampling across encoder layers, improving separable speaker representation at every conformer layers. \textbf{(iii)} We conduct a systematic ablation study to identify the architectural components most beneficial for self-supervisesd multi-speaker representations. \textbf{(iv)} We evaluate SepRQ on a broad set of multi-speaker downstream tasks, out-domain datasets and challenging three speakers mixtures, to evaluate the scalability and generalizibility of our approach. \textbf{(v)} Finally, we make our SepRQ model available \footnote{SepRQ available at : \url{https://sevkod.github.io/SepRQ/}} to the community in order to help promote future research on the topic of separation-based and multi-speaker SSL.

\section{SepRQ}
\label{sec:seprq}

Following the initial approach introduced by C-HuBERT, our objective is to enforce source separation directly in a shared discrete representation space by predicting a separate sequence of quantized units for each speaker, as illustrated in Figure~\ref{fig:seprq-multires}. Rather than explicitly reconstructing the separated waveforms, the model is trained to disentangle the discrete representations corresponding to the individual speakers present in a mixture.



Given a target number of $K$ output sources, we construct each training mixture by sampling up to $K$ single-speaker utterances $\{x_k\}$, with a decreasing $50\%$ probability of adding 1 source, 2 source, \ldots K sources. Each utterance is independently scaled by an uniformly sampled gain $g_k$ to account for variations in relative speaker loudness. The decreasing probability of including additional sources reflects the lower presence of real-world mixtures having three or more overlapping speakers.

As demonstrated by WavLM, a denoising pretext task is essential to produce robust speech representations in noisy scenarios. Since noise is a fundamental aspect of cocktail-party problems, we add a background noise utterance $x_b$ and a low-gain interfering speech utterance $x_n$ on top of the mixture, scaled uniformly by $g_b$ and $g_n$ respectively, each being added independently with a probability of $20\%$, to promote overall variability in generated interferences.
The resulting mixture $x_{\mathrm{mix}}$ is defined as:
\begin{equation}
x_{\mathrm{mix}}
=
g_b x_b
+
g_n x_n
+
\sum_{k=1}^{K} g_k x_k .
\end{equation} 




In a cocktail-party setting, we want to enforce the mixture and its clean constituent speakers to lie in the same representation space, despite differences in background noise and speaker gain. This ensures that acoustic and semantic information remain meaningful across the sources to disentangle. We therefore normalize both mixture and clean-reference spectrograms with a shared global mean and standard deviation, estimated over a portion of the mixture and clean reference training data via moving average. 

The normalized mixture spectrogram is processed by two convolutional layers followed by twelve Conformer layers, following the core architecture of BestRQ \cite{bestrq}. BestRQ's original front-end operates at 25~Hz, a frame rate that may be too coarse for cocktail-party generation tasks such as separation or enhancement, which need fine-grained representations for accurate reconstruction. We therefore halve the stride of the final convolutional layer, raising the frame rate to 50~Hz (20~ms) and matching WavLM and C-HuBERT for direct comparison.

The central contribution of SepRQ is the introduction of separation objectives at multiple temporal resolutions. A separation objective is applied after every two consecutive Conformer layers, with progressive temporal downsampling so that the entire encoder captures both fine-grained acoustics and longer-term structure. The downsampling process folds (or concatenate) $H$ consecutive latent frames along the hidden dimension (a factor-of-two for every additional pair of Conformer layers), gradually reducing the resolution (for 12 layers, this represents $H=6$ foldings : 20~ms $\rightarrow$ 40~ms  $\rightarrow$ 80~ms $\rightarrow$ 160~ms $\rightarrow$ 320~ms). Through experimentation, folding has shown to outperform methods such as mean or attention pooling. 

While WavLM relies on $k$-means clustering to construct the discrete targets used for the masked prediction objective, we instead employ the more practical frozen Random Vector Quantizer (RVQ) approach introduced in the BestRQ framework to map our clean sources onto a shared discrete codebook. In a multiresolution setting, we assign a dedicated codebook to each temporal scale to account for the differences in structured information captured at each scale.
Each separation layer uses $K$ independent projection heads, which can be either linear layers or multilayer perceptrons (MLPs). Since folding frames at a specific layer essentially multiplies the hidden dimension by a scaling factor $H$, the overall parameter count grows exponentially as we add more layers/downsampling operations. To compensate we halve the codebook dimensionality at each downsampling stage (8192 codewords at 50 Hz, 4096 at 25 Hz, and proportionally adjusted for other sampling rates), keeping the pre-training parameter count manageable. 
Lowering the frame rate reduces the number of speech representations that need to be quantized. Empirically, we find that these lower-rate representations can be effectively quantized using a smaller codebook, without degrading downstream performance.

As speaker assignment has no predefinite order for each set of heads, we apply utterance-level Permutation Invariant Training (PIT) independently at each resolution. Since each head has its specific set of features, temporal fold, and codebook, we hypothesize that the best speaker-to-stream match is local to that head. At inference time, only the encoder layers are retained, therefore avoiding the need to solve any permutation problem. For codebook $h$, the pseudo source-separation loss is the utterance-level permutation-invariant cross-entropy, averaged over the $K$ sources and $T_h$ frames at that resolution:
\begin{equation}
\mathcal{L}_{\mathrm{PSS}}^{(h)}
=
\min_{\pi \in \mathcal{P}_K}
\frac{1}{K T_h}
\sum_{i=1}^{K}
\sum_{\tau=1}^{T_h}
CE\!\left(t_{i,\tau}^{(h)}, p_{\pi(i),\tau}^{(h)}\right)
\end{equation}
where $\mathcal{P}_K$ is the set of permutations of the $K$ sources, $t_{i,\tau}^{(h)}$ and $p_{j,\tau}^{(h)}$ are the target and predicted codewords at frame $\tau$, and $CE$ denotes the per-frame cross-entropy. 
The multiresolution objective averages these losses over the $H$ prediction layers:
\begin{equation}
\mathcal{L}_{\mathrm{MR\text{-}PSS}}
=
\frac{1}{H}
\sum_{h=1}^{H}
\mathcal{L}_{\mathrm{PSS}}^{(h)}
\end{equation}
Finally, unlike C-HuBERT, we do not employ an additional masked modeling objective on top of the PSS objective. Without masking, the model predicts discrete units across the entire utterance rather than only at masked positions, making more efficient use of the training data by leveraging the full sequence for prediction. The absence of masking also enables our multiresolution approach to work in a SSL fashion, as otherwise, masked frames could be folded together with unmasked ones.
\section{EXPERIMENTS}

\begin{table}[t]
  \centering
  \caption{Ablation study of SepRQ design choices on SUPERB speech separation (SI-SDRi, dB) and speaker diarization (DER, \%)}
  \label{tab:pretext-ablation-sep}
  \scriptsize
  \setlength{\tabcolsep}{4pt}
  \begin{tabular}{c|l|r|r}
  \toprule
  \textbf{Dataset} & \textbf{Model} & \textbf{SI-SDRi} $\uparrow$ & \textbf{DER} $\downarrow$ \\
  \midrule
  \multirow{8}{*}{\textit{LS-100h}}
    & BestRQ (50~Hz) & $9.55$ & $7.24$ \\
    & $+$ Trained on mixtures & $10.03$ & $5.91$ \\
    & \hspace{1em}$+$ $\mathcal{L}_{\mathrm{PSS}}^{(6)}$ only & $10.57$ & $4.01$ \\ [0.07cm]
    & \hspace{2em}$-$ Masked Language Modeling & $10.87$ & $3.57$ \\
    & \hspace{3em}$+$ Multilayer Perceptron & $10.81$ & $3.42$ \\
    & \hspace{4em}$+$ Multiresolution (\textbf{SepRQ}) & $\mathbf{11.22}$ & $\mathbf{2.73}$ \\
  \midrule
\multirow{2}{*}{\textit{LS-960h}}
  & \hspace{3em}$+$ Multilayer Perceptron & $11.51$ & $2.35$ \\
  & \hspace{4em}$+$ Multiresolution (\textbf{SepRQ}) & $\mathbf{12.10}$ & $\mathbf{2.08}$ \\
  \bottomrule
  \end{tabular}
\end{table}


\newcommand{\cmark}{\ding{51}}
\newcommand{\xmark}{\ding{55}}

Our experimental protocol proceeds in two stages: first, we perform a reduced-budget ablation on LibriSpeech-100~h (single H100 GPU) to isolate the contribution of each SepRQ design choice by scoring every variant on the SUPERB speech separation and diarization~\cite{superb} tasks; the selected configuration is then scaled to full pre-training and evaluated under both in-domain multi-speaker benchmarks and out-of-domain (OOD) conditions. For full scale, we pretrain two SepRQ models with $K{=}2$ ($3$ resp.) output sources using a 12-layer Conformer encoder for $266.9$M ($357.5$M resp.) parameters; ${\approx}41$ (${\approx}49$ resp.) GFLOPs per second of audio, on mixtures assembled on the fly from LibriSpeech~960~h ~\cite{librispeech} utterances. The mixtures are optionally augmented with WHAM~\cite{wham} noise and LibriSpeech interfering utterances at SNRs up to $15$~dB, as described in section \ref{sec:seprq}. We train all full scales on four H100 GPUs with batch sizes of $1000$ ($500$ resp.) seconds of audio per GPU, and gradient accumulation factors of $2$ ($4$ resp.). The learning rate follows a trapezoidal schedule (linear warmup over $16{,}000$ steps, constant rate $1.4{\times}10^{-3}$ until step $150{,}000$, linear cooldown over the final $30{,}000$ steps; $180{,}000$ steps total). We experimentally observe a noticeable improvement in performance using a constant learning rate for a long period of time, as the model is equally exposed to a diverse combination of possible mixtures.

At inference time, we freeze SepRQ and retain only the convolutional front-end, Conformer encoder, and global normalization statistics, dropping the total effective parameter count down to $85.68$M parameters (${\approx}8.7$~GFLOPs per second of audio at $50$~Hz). As frame folding is used only during pre-training as a pretext task for the separation objective, the multi-speaker structural information remains encoded in the Conformer representation, allowing inference to run at $50$~Hz as standard SSL models would. All our downstream tasks use a weighted-average representation of the twelve Conformer outputs.

We first evaluate SepRQ on the multi-speaker SUPERB suite, using for all following tasks the hyperparameters used in \cite{WavLM}: Speaker Diarization (SD), Speech separation (SS), and Speech Enhancement (SE). We then evaluate the scalability of SepRQ accross other tasks, on the target-speaker benchmark TS-SUPERB. The latter includes Target Speaker Extraction (TSE), Personalized Speaker Extraction (PSE), Personalized Voice Activity Detection (PVAD), and TS-ASR, all trained on Libri2Mix (for TSE and TS-ASR) / SparseLibriMix~\cite{librimix} (for PVAD and PSE). All chosen hyperparameters follow the implementation from~\cite{ts_superb}. Finally, to probe OOD robustness of SepRQ, we evaluate EEND-based diarization \cite{eend} on DIHARD~3~\cite{dihard3} and ConvTasNet \cite{convtasnet} separation on WSJ0-2/3mix~\cite{wsj0}, following the protocol and hyperparameters described in~\cite{asru25}. For SS, SE, TSE, and PSE, we either report the Scale-Invariant Signal-to-Distortion Ratio Improvement (SI-SDRi), SDRi or Short-Time Objective Intelligibility (STOI). For SD and PVAD, we report Diarization Error Rate (DER) and the mean Average Precision (mAP) of the target speaker respectively. TS-ASR is evaluated using the Word Error Rate (WER) of the extracted target-speaker.

\begin{table}[t]
  \caption{Multi-speaker SUPERB results (speaker diarization, speech separation, and speech enhancement) for SepRQ. Pre-training data in hours: LS = LibriSpeech, LL = Libri-Light, Mix = LL-60k + GigaSpeech-10k + VoxPopuli-24k}
  \label{tab:superb_metrics}
  \centering
  \footnotesize
  \setlength{\tabcolsep}{3pt}
  \resizebox{\columnwidth}{!}{%
  \begin{tabular}{l | c | c | c | c | c c}
    \toprule
    \multirow{2}{*}[-1pt]{\textbf{Methods}} &
    \multirow{2}{*}[-2pt]{\begin{tabular}[c]{@{}c@{}}\textbf{\#Param}\\\textbf{(M)}\end{tabular}} &
    \multirow{2}{*}[-2pt]{\begin{tabular}[c]{@{}c@{}}\textbf{Data}\\\textbf{(h)}\end{tabular}} &
    \textbf{SD} & \textbf{SS} & \multicolumn{2}{c}{\textbf{SE}} \\
    \cmidrule(lr){4-4} \cmidrule(lr){5-5} \cmidrule(lr){6-7}
    & & & \textbf{DER}$\downarrow$ & \textbf{SI-SDRi}$\uparrow$ & \textbf{PESQ}$\uparrow$ & \textbf{STOI}$\uparrow$ \\
    \midrule
    \multicolumn{7}{l}{\textit{Base-class} (${\approx}95$M)} \\
    \midrule
    \rowcolor{gray!15}
    \multicolumn{7}{l}{\textit{Generic SSL Models}} \\
    HuBERT Base    & 94.68 & LS-960  & 5.88 & 9.36  & 2.58 & 93.9 \\
    WavLM Base     & 94.70 & LS-960  & 4.55 & 10.37 & 2.58 & 94.0 \\
    WavLM Base+    & 94.70 & Mix-94k & 3.50 & 10.85 & 2.63 & 94.3 \\
    \rowcolor{gray!15}
    \multicolumn{7}{l}{\textit{Cocktail-Party SSL Models}} \\
    C-HuBERT Base  & 96.00 & LS-960  & 2.77 & 11.08 & 2.63 & 94.0 \\
    \textcolor{gray!70!black}{SA-WavLM$^\dagger$} & \textcolor{gray!70!black}{94.97} & \textcolor{gray!70!black}{LS-960} & \textcolor{gray!70!black}{1.88} & \textcolor{gray!70!black}{11.13} & \textcolor{gray!70!black}{2.62} & \textcolor{gray!70!black}{94.2} \\
    \textbf{SepRQ (ours)} & \textbf{85.68} & \textbf{LS-960} & \textbf{2.08} & \textbf{12.10} & \textbf{2.67} & \textbf{94.4} \\
    \midrule
    \multicolumn{7}{l}{\textit{Large-class} (${>}316$M)} \\
    \midrule
    \rowcolor{gray!15}
    \multicolumn{7}{l}{\textit{Generic SSL Models}} \\
    HuBERT Large   & 316.62 & LL-60k  & 5.75 & 10.45 & 2.64 & 94.2 \\
    WavLM Large    & 316.62 & Mix-94k & 3.24 & 11.19 & \underline{2.70} & \underline{94.5} \\
    \rowcolor{gray!15}
    \multicolumn{7}{l}{\textit{Cocktail-Party SSL Models}} \\
    C-HuBERT Large & 318.00 & LL-60k  & 2.65 & 11.24 & 2.65 & 94.3 \\
    \bottomrule
  \end{tabular}%
  }
\end{table}

\subsection{Ablation study}

Table~\ref{tab:pretext-ablation-sep} isolates each core component of SepRQ, comparing each architectural choice with the SS and SD downstream tasks of SUPERB. One could argue that the improvements on multi-speaker tasks are simply a consequence of domain matching: since SepRQ is pretrained on mixtures rather than single-speaker audio, the encoder may benefit from seeing overlapped speech, independently of the separation-oriented objective.
To address this, we train a reduced-budget BestRQ model on the same on-the-fly mixture process used by SepRQ, comparing it to the pseudo source separation approach applied at the final encoder layer only ($\mathcal{L}_{\mathrm{PSS}}^{(6)}$).
A cocktail-party loss improves SS by $0.54$~dB SI-SDRi over mixture-trained BestRQ, confirming a real benefit from a PSS objective, rather than from simply applying Masked Language Modelling to mixture audio.
Removing masking for full-utterance separation further raises SI-SDRi, showing that masking indeed bottlenecks separation performance.
Finally, although MLPs slightly underperform linear heads in terms of SI-SDRi, they achieve more accurate codeword prediction during pretraining, which translates into improved downstream diarization performance. Furthermore, we observe that MLP heads yield better separation results when combined with the multiresolution setup, because projectors are deployed across multiple layers.
At full scale, the multiresolution approach yields a more substantial gain in separation than under the reduced budget ($+0.59$~dB SI-SDRi), displaying both the scalability and the transferability in gains of SepRQ from low to high budget.


\subsection{SUPERB \& TS-SUPERB}


Table~\ref{tab:superb_metrics} summarizes the results of the SepRQ model on the multi-speaker SUPERB suite, compared to generic and cocktail-party SSL baselines, for Base and Large versions. 
\begingroup
\makeatletter
\def\@makefnmark{\hbox{\normalfont$^{\dagger}$}}%
\footnotetext{Uses oracle speaker embeddings of the target speakers at inference time.}%
\endgroup
SepRQ sets a new state-of-the-art on diarization and separation among enrollment-free SSL models, beating Base and Large versions, with the fewest parameters (85.68M) and pretraining data. Against WavLM Base, SepRQ reduces DER by 54.3\% relative and improves SI-SDRi by an absolute $+1.7$~dB. It also outperforms C-HuBERT, reducing DER by 24.9\% over its Base version, and improving SI-SDRi by a noticeable $+1.02$~dB. The model also achieves state-of-the-art performance on SE accross all the base-class models, beating WavLM Base and Base+, while also outperforming C-HuBERT Large and remaining competitive with WavLM Large.
Compared to SA-WavLM, SepRQ improves SI-SDRi in SS by a similar margin ($+0.97$~dB), while also achieving higher PESQ and STOI in SE. On SD, SA-WavLM retains a slight edge (1.88 vs. 2.08 DER); however SA-WavLM exploits prior speaker identity information during pretraining, an advantage directly aligned with diarization. SepRQ, in contrast, remains fully enrollment-free.

\begin{table}[ht]
  \caption{TS-SUPERB results for SepRQ on target speech extraction (TSE), personalized speech extraction (PSE), personalized VAD (PVAD), and target-speaker ASR (TS-ASR).}
  \label{tab:audio_metrics}
  \centering
  \footnotesize
  \setlength{\tabcolsep}{2.5pt}
  \resizebox{\columnwidth}{!}{%
  \begin{tabular}{l | c c | c c | c | c c}
    \toprule
    \multirow{2}{*}[-1pt]{\textbf{Upstream}} & \multicolumn{2}{c|}{\textbf{TSE}} & \multicolumn{2}{c|}{\textbf{PSE}} & \textbf{PVAD} & \multicolumn{2}{c}{\textbf{TS-ASR} (\textbf{WER}$\downarrow$)} \\
    \cmidrule(lr){2-3} \cmidrule(lr){4-5} \cmidrule(lr){6-6} \cmidrule(lr){7-8}
    & \textbf{SI-SDRi}$\uparrow$ & \textbf{STOI}$\uparrow$ & \textbf{SI-SDRi}$\uparrow$ & \textbf{STOI}$\uparrow$ & \textbf{mAP}$\uparrow$ & w/o LM & w/ LM \\
    \midrule
    HuBERT Base  & 9.64  & 87.30 & 8.61  & 79.92 & 94.60 & 36.86 & 30.52 \\
    WavLM Base   & 10.26 & 88.40 & 9.65  & 81.57 & 94.40 & 27.82 & 22.68 \\
    WavLM Base+  & 10.69 & 89.00 & 10.01 & 82.67 & 95.00 & 24.75 & 20.06 \\
    \midrule
    SepRQ & \textbf{12.53} & \textbf{91.45} & \textbf{11.18} & \textbf{85.60} & \textbf{96.61} & \textbf{22.26} & \textbf{16.78} \\
    \bottomrule
  \end{tabular}%
  }
\end{table}
On TS-SUPERB (Table~\ref{tab:audio_metrics}), SepRQ is best across all current SSL models reported to date on TSE, PSE, PVAD, and TS-ASR.
Relative to WavLM Base+, extraction improves by $+1.84$~dB on TSE and $+1.17$~dB on PSE.
The same trend appears on TS-ASR, displaying a relative improvement of around 5\% in WER over WavLM Base, with or without using a Language Model (LM) for decoding. The lower WER achieved with SepRQ indicates not only improved overall signal reconstruction but enhanced extraction of the semantic of the target-speaker, consistent with the higher STOI scores observed for both TSE and PSE. This gap becomes even more pronounced when using a language-model (LM) head for ASR, when comparing to other SSL models.
Finally, at the dataset scale, SepRQ outperforms WavLM Base+ and Large on every reported SUPERB and TS-SUPERB tasks despite requiring a much smaller corpus ($960$~h vs.\ ${\approx}94$k~h), proving the effectiveness of the on-the-fly mixture pre-training process.

\begin{table}[ht]
  \caption{Out-of-domain evaluation on DIHARD~3 diarization (FA/MD/SC: False Alarm / Missed Detection / Speaker Confusion; DER: Diarization Error Rate) and WSJ0-2mix / WSJ0-3mix separation SDRi (Signal-to-Distortion Ratio improvement)}
  \label{tab:ood_performance}
  \centering
  \footnotesize
  \setlength{\tabcolsep}{4pt}
  \renewcommand{\arraystretch}{1.05}
  \resizebox{\columnwidth}{!}{%
  \begin{tabular}{l cc | cc}
    \toprule
    \multirow{3}{*}{\textbf{Methods}} & \multicolumn{2}{c}{\textbf{Diarization}} & \multicolumn{2}{c}{\textbf{Separation}} \\
    \cmidrule(lr){2-3} \cmidrule(lr){4-5}
    & \multicolumn{2}{c}{\textbf{DIHARD 3}} & \textbf{WSJ0-2mix} & \textbf{WSJ0-3mix} \\
    \cmidrule(lr){2-3} \cmidrule(lr){4-4} \cmidrule(lr){5-5}
    & \textbf{FA/MD/SC \%}$\downarrow$ & \textbf{DER}$\downarrow$ & \textbf{SDRi}$\uparrow$ & \textbf{SDRi}$\uparrow$ \\
    \midrule
    w/o SSL features & 6.1 / 8.1 / 4.9 & 19.3 & 16.4 & 13.1 \\
    \midrule
    HuBERT Base     & 4.3 / 8.6 / 4.7 & 17.7 & 17.0 & 12.7 \\
    WavLM Base      & 4.4 / 8.3 / 4.5 & 17.3 & 17.5 & 13.2 \\
    WavLM Base+ & 4.8 / 7.6 / 4.2 & 16.6 & 18.2 & 13.4 \\
    \midrule
    SepRQ ($2 \; sources$) & 4.9 / 7.4 / 3.8 & \textbf{16.1} & \textbf{19.8} & 14.5 \\
    SepRQ ($3 \; sources$) & 5.0 / 7.3 / 4.1 & 16.4 & 19.4 & \textbf{17.2} \\
    \bottomrule
  \end{tabular}%
  }
  \vspace{-1em}
\end{table}

\subsection{Out-of-domain evaluation}

On OOD evaluation (Table~\ref{tab:ood_performance}), SepRQ outperforms WavLM Base and the other SSL baselines on DIHARD~3, confirming usefulness in realistic conversational conditions through lower missed-detection and speaker-confusion rates. The $K{=}2$ model slightly outperforms $K{=}3$ on DER, which we attribute to mild over-specialization of the three-sources model toward three-speaker mixtures. On WSJ0-2mix, SepRQ ($K{=}2$) improves SDRi by $2.3$~dB over WavLM Base and $3.4$~dB over a system with no SSL features, supporting its use as a strong upstream SSL model for speech separation.
The gain is more pronounced on WSJ0-3mix: while WavLM Base and even Base+ struggle to improve the SDRi over the no-SSL baseline, SepRQ trained on $K{=}2$ sources already gains $1.4$~dB of SDR improvement. The $K{=}3$ source version nearly triples that margin ($+4.1$~dB over WavLM Base); displaying strong separation capabilities in more challenging cocktail-party scenarios, while retaining stable performance on WSJ0-2mix, and DIHARD~3.

\section{CONCLUSION}
\label{sec:format}

We present SepRQ, an open-source self-supervised framework that learns mixture-aware representations through multiresolution pseudo-source separation. 
Thanks to an elaborate synthetic mixture generation process and a mask-free approach, SepRQ adopts a discrete multi-scale source separation objective to enforce linearly separable speaker representations at multiple levels. SepRQ achieves state-of-the-art performance on multi-speaker SUPERB and all reported TS-SUPERB tasks, beating mainstream models such as WavLM, or closed-source cocktail-party approaches such as C-HuBERT. Our approach transfers effectively to out-of-domain DIHARD~3 diarization and WSJ0-2mix~/~WSJ0-3mix separation, with particularly strong gains on three-speaker mixtures. Despite its strong performance on multi-speaker tasks, a non-finetuned SepRQ remains limited on single-speaker tasks such as ASR. We hypothesize that combining a single speaker pretraining modality with the current PSS approach, or adopting unified frameworks~\cite{unifying} are promising directions to reduce this gap.

\section{Acknowledgments}
This work was granted access to the HPC resources of IDRIS under the allocations AD011014044R2, 2024-AD011015163 and AD011016519 made by GENCI, and supported by the Agence de l'Innovation Defense under the grant number 2022 65 0079. It also benefited from the support of the French National Research Agency through the ANR-20-CE23-0012-01 (MIM) grant. LLMs (Gemini 3 Pro, ChatGPT) were used only to edit certain portions of the manuscript's wording.

\section{Compliance with Ethical Standards}
This research was conducted retrospectively using human subject speech data made available in open access or under institutional license (LibriSpeech, WHAM!, WSJ0, DIHARD 3). Ethical approval was not required, as confirmed by the licenses of these corpora. No new data were collected from human subjects.
\bibliographystyle{IEEEbib}
\bibliography{ref2}

\end{document}